%% file: main.tex
\documentclass[conference]{IEEEtran}
\IEEEoverridecommandlockouts

\usepackage{cite}
\usepackage{amsmath,amssymb,amsfonts}
\usepackage{algorithmic}
\usepackage{graphicx}
\usepackage{textcomp}
\usepackage[dvipsnames, table]{xcolor}
\usepackage{paralist}
\usepackage[inline]{enumitem}
\usepackage{caption}
\usepackage{makecell}
\usepackage{tabu, booktabs}
\usepackage{float}
\usepackage{subcaption}
\usepackage{comment}
\usepackage[breaklinks,colorlinks,bookmarks=true]{hyperref}
\usepackage{csquotes}
\usepackage{soul}
\usepackage{pifont}
\usepackage{multirow}
\usepackage{tcolorbox}
\usepackage{tabularx}
\usepackage{array}
\usepackage{colortbl}
\usepackage{siunitx}
\usepackage{fancyhdr}

\definecolor{comp_pos}{HTML}{63B438}
\definecolor{comp_neg}{HTML}{E1564D}
\definecolor{comp_neut}{HTML}{A9AEA6}

\definecolor{magma_darker}{HTML}{fdc38a}
\definecolor{magma_dark}{HTML}{e15666}
\definecolor{magma_light}{HTML}{82247f}
\definecolor{magma_lighter}{HTML}{1f0c43}

\definecolor[named]{xBlue}{HTML}{18647E}
\definecolor[named]{xOrange}{HTML}{FF9B00}
\definecolor[named]{xGray}{HTML}{808080}
\definecolor[named]{xGreen}{HTML}{60B950}
\definecolor[named]{xRed}{HTML}{A30B37}
\definecolor[named]{xDarkBlue}{cmyk}{1,0.58,0,0.21}

\definecolor{cadetblue}{rgb}{0.37, 0.62, 0.63} % A variant of cadet blue
\definecolor{grayish}{rgb}{0.6, 0.6, 0.6} % A variant of grey
\definecolor{redorange}{rgb}{0.91, 0.41, 0.17} % A variant of red orange
\definecolor{limegreen}{rgb}{0.5, 0.8, 0.2} % A variant of lime green

\hypersetup{
    colorlinks=true,
    linkcolor=xOrange,
    filecolor=magenta,      
    urlcolor=xDarkBlue,
    citecolor=xBlue,
}

\usepackage[a4paper, total={184mm,239mm}]{geometry}
\def\BibTeX{{\rm B\kern-.05em{\sc i\kern-.025em b}\kern-.08em
    T\kern-.1667em\lower.7ex\hbox{E}\kern-.125emX}}
    
\ExplSyntaxOn
\DeclareExpandableDocumentCommand{\convertlen}{ O{cm} m }
 {
  \dim_to_decimal_in_unit:nn { #2 } { 1 #1 } cm
 }
\ExplSyntaxOff
\usepackage{gnuplottex}
\usepackage{tikz}
\usepackage{mathtools}

\usetikzlibrary{matrix,calc,positioning,arrows.meta}

\newcommand{\cmark}{\textcolor{green!70!black}{\ding{51}}}
\newcommand{\xmark}{\textcolor{red}{\ding{55}}}

\newcommand*\circled[1]{\tikz[baseline=(char.base)]{\node[shape=circle,fill,inner sep=0.5pt] (char) {\textcolor{white}{#1}};}}

\usepackage{tcolorbox}
\usepackage[normalem]{ulem}

\begin{document}

\title{Less is More: Optimizing Function Calling for LLM Execution on Edge Devices}

\author{
\IEEEauthorblockN{Varatheepan~Paramanayakam$^{1}$, Andreas Karatzas$^{1}$, Iraklis Anagnostopoulos$^1$, Dimitrios Stamoulis$^2$}
\IEEEauthorblockA{$^1$School of Electrical, Computer and Biomedical Engineering, Southern Illinois University, Carbondale, IL, U.S.A.}
\IEEEauthorblockA{$^2$Department of Electrical and Computer Engineering, The University of Texas at Austin, Austin, TX, U.S.A.}
\IEEEauthorblockA{Email: \{varatheepan, andreas.karatzas, iraklis.anagno\}@siu.edu, dstamoulis@utexas.edu}
}

\maketitle
\input{abstract}

\thispagestyle{fancy}
\fancyhead{}
\rhead{}
\lhead{}
\chead{{\small Accepted for publication at the 28th Design Automation and Test in Europe Conference (DATE 2025)}}
\renewcommand{\headrulewidth}{0.4pt}


\setlist{nosep}

\input{introduction}

\input{related}
\input{methodology}
\input{evaluation}
\input{conclusion}
\input{acknowledgment}

\clearpage
\bibliographystyle{IEEEtran}
\footnotesize
\bibliography{ref}
\end{document}

%% file: abstract.tex
\begin{abstract}
The advanced function-calling capabilities of foundation models open up new possibilities for deploying agents to perform complex API tasks. However, managing large amounts of data and interacting with numerous APIs makes function calling hardware-intensive and costly, especially on edge devices. Current Large Language Models (LLMs) struggle with function calling at the edge because they cannot handle complex inputs or manage multiple tools effectively. This results in low task-completion accuracy, increased delays, and higher power consumption. In this work, we introduce Less-is-More, a novel fine-tuning-free function-calling scheme for dynamic tool selection. Our approach is based on the key insight that selectively reducing the number of tools available to LLMs significantly improves their function-calling performance, execution time, and power efficiency on edge devices. Experimental results with state-of-the-art LLMs on edge hardware show agentic success rate improvements, with execution time reduced by up to 70\% and power consumption by up to 40\%.
\end{abstract}

%% file: introduction.tex
\section{Introduction and motivation}\label{sec:intro}

Among the numerous advances in Generative AI, a powerful emerging paradigm is leveraging the reasoning capabilities of Large Language Models (LLMs) within agentic systems that can execute the appropriate API tools to complete user queries~\cite{patil2023gorilla}. However, as function calling involves complex tasks, data handling, and API interactions, deploying such agents introduces significant hardware bottlenecks~\cite{karatzas2024pythia,yuan2024llm}.

Recent approaches from both industry and academia -- such as OpenAI's parallel tool-calling capabilities~\cite{openai2024fun} and novel prompting techniques~\cite{yao2023react} -- have contributed to system-efficient function calling. Unfortunately, most solutions focus on serverless GPT-based deployments, severely limiting their applicability on the edge~\cite{karatzas2024pythia,stamoulis2019single}. As highlighted by recent announcements from smart-device providers like Apple, there is a growing need for efficient on-device execution, especially in applications where user data cannot be sent to the cloud~\cite{abdin2024phi}.

A common approach is to replace large billion-parameter models (e.g., 70B, 400B) with \enquote{\textit{smaller}} LLMs that have fewer than 10B parameters (e.g., 1.5B, 3.8B, 7B, 8B)~\cite{abdin2024phi}. Even with fewer parameters, the default versions of these models still have significant overheads in terms of execution time and power consumption~\cite{yuan2024llm}. Libraries like HuggingFace~\cite{hf-tool} and Ollama~\cite{ollama-tool} offer optimizations such as aggressive quantization, token pruning, and context window reduction to improve efficiency on edge hardware. However, these smaller, quantized/pruned LLMs often lack sophisticated function-calling abilities~\cite{erdogan2024tinyagent} and require costly refinement and fine-tuning to reach the agentic performance of larger models~\cite{abdin2024phi}. Notably, when deploying ``off-the-shelf'' open-source LLMs \ul{\textbf{without}} additional fine-tuning, there is a significant performance drop. %For instance, 
As shown in Table~\ref{table:motivation_success_rate}, the success rates on the Berkeley Function-Calling Leaderboard (BFCL)~\cite{berkeley-function-calling-leaderboard} and GeoEngine~\cite{singh2024geollm} benchmarks%\footnote{More information about experimental setup can be found in Section~\ref{sec:evaluation}} 
drop substantially when comparing quantized/pruned LLMs to their full-precision HuggingFace counterparts.

\begin{table}
\centering
\caption{Success rate of Llama3.1-8b variations~\cite{ollama-tool} on the BFCL~\cite{berkeley-function-calling-leaderboard} and GeoEngine~\cite{singh2024geollm} benchmarks.}
% \small
\label{table:motivation_success_rate}
\resizebox{0.95\columnwidth}{!} 
{ 
\begin{tabular}{@{}cccccc@{}}
\toprule
\textbf{Benchmark} & \textbf{Full precision} & \textbf{q4\_0} & \textbf{q4\_1} & \textbf{q4\_K\_M} & \textbf{q8\_0} \\
\midrule
BFCL~\cite{berkeley-function-calling-leaderboard} & 63.04\% & 20.43\% & 34.35\% & 39.57\% & 44.35\% \\
GeoEngine~\cite{singh2024geollm} & 63.91\% & 43.04\% & 59.57\% & 56.96\% & 53.04\% \\
\bottomrule
\end{tabular}
}
\vspace*{-4pt}
\end{table}

\begin{table}
\centering
\caption{Execution of function-calling query using Llama3.1-8b-q4\_K\_M~\cite{ollama-tool} on Nvidia Jetson AGX Orin~\cite{karumbunathan2022nvidia}.}
\scriptsize
\label{table:motivation_exec_time}
\resizebox{0.95\columnwidth}{!} 
{ 
\begin{tabular}{@{}ccccc@{}}
\toprule
\textbf{Context window} & \textbf{\# Tools} & \textbf{Successful} & \textbf{Exec. time (s)} & \textbf{Power (W)} \\
\midrule
16K & 46 & \xmark & 30 & 27 \\
16K & 19 & \cmark & 20 & 26 \\
8K & 19 & \cmark & 17 & 22 \\
\midrule
Max drop & & & \textcolor{comp_pos}{$\downarrow 43\%$} & \textcolor{comp_pos}{$\downarrow 19\%$} \\
\bottomrule
\end{tabular}
}
\vspace*{-12pt}
\end{table}

Furthermore, we would like to illustrate that enabling function calling on edge devices to support advanced functionality is a complex task, often resulting in high execution time and power consumption. Take, for example, the following query:
% \vspace{-20pt}
\begin{tcolorbox}[
  colback=cadetblue!5, % Light tint of cadet blue for content
  colframe=cadetblue,
  colbacktitle=cadetblue!20,
  coltitle=black,
  title=Query example from GeoEngine benchmark~\cite{singh2024geollm},
  fonttitle=\bfseries
  % ~\footnote{From GeoEngine~\cite{singh2024geollm} benchmark. More information in Section~\ref{sec:evaluation}}
]
\vspace{-6pt}
\emph{Plot the fmow VQA captions in UK from Fall 2009}
\vspace{-4pt}
\end{tcolorbox}
\vspace{-4pt}
\noindent GeoEngine provides 46 tools with each query, allowing the LLM to decide which ones to use. As shown in Table~\ref{table:motivation_exec_time}, even though Llama3.1-8b-q4\_K\_M has a 16K context window that can fit all the tools, it fails to select the correct one. This occurs because of the large number of available options confusing the LLM. However, by reducing the number of tools provided, the model's reasoning ability improves. For instance, when only 19 tools are passed, the LLM chooses the correct tool and completes the query successfully. This reduction in tools helps the LLM focus and results in a higher success rate. Moreover, it significantly decreases execution time. In essence, providing \ul{\textbf{fewer}} options enables the LLM to make more accurate and faster decisions. Therefore, it is crucial to develop a mechanism that \emph{dynamically reduces the set of tools available to the LLM while adjusting the context window accordingly to optimize both performance and energy efficiency on edge devices}.

In this work, we introduce \texttt{Less-is-More}, a novel fine-tuning-free function-calling scheme for dynamic
tool selection. Our \textbf{key insight} is that selectively reducing the number of tools available to the LLM significantly improves its decision-making ability. By presenting the LLM with fewer, \emph{more relevant tools}, we reduce confusion, allowing the model to focus better and achieve higher accuracy.
Additionally, this reduction improves execution time and power consumption, and it allows the use of smaller context windows, enhancing both execution speed and energy efficiency further. \texttt{Less-is-More} achieves these improvements without requiring any model retraining or fine-tuning, making it a practical solution for optimizing LLM performance on resource-constrained edge devices.

\textbf{Overall, our main contributions are:}
\begin{inparaenum}
    \item[\circled{1}] We introduce \texttt{Less-is-More}, a new approach for dynamic tool selection that optimizes function-calling in LLMs without requiring any finetuning or retraining, making it a plug-and-play solution for all existing state-of-the-art LLMs.
    \item[\circled{2}] Our method boosts LLM task-completion success rates by reducing the number of tools provided to the LLM, which reduces tool-space complexity and enables the agent to make more accurate and efficient decisions during function calls.
    \item[\circled{3}] We achieve significant reductions in power consumption and response time for LLMs running on edge devices, improving overall performance and efficiency.
\end{inparaenum}

%% file: related.tex
\section{Related Work}\label{sec:related}

Many recent approaches aim to improve the runtime efficiency of LLMs, such as quantization~\cite{dettmers2024qlora}, pruning~\cite{ma2023llmpruner}, KV token caching~\cite{kwon2023kv}, knowledge distillation~\cite{gu2024minillm}, token compression~\cite{jiang2023llmlingua}, in-memory execution~\cite{alizadeh2023llmapple,kim2024llmmemory}, and speculative decoding~\cite{miao2024specinfer}. However, existing methods are evaluated primarily in GPT-style conversational flows or text generation tasks, limiting their applicability to function calling~\cite{li2024llmsurvey}, where hardware-efficient deployment remains a bottleneck. As we demonstrated in Section~\ref{sec:intro}, when quantized models with pruned window sizes are used as agents, there is a significant performance drop. In contrast, our approach focuses on enhancing agentic capabilities and maintaining performance consistency across different models, whether quantized, fine-tuned, or otherwise.

Edge-LLM~\cite{yu2024edgellm} employs adaptive layer voting for efficient execution on edge devices. Recent work extends these voting schemes to Mixture-of-Experts~\cite{hu2024routerbench} and LLM cascading~\cite{chen2023frugalgpt}, dynamically selecting between more powerful and weaker models based on a local-remote paradigm~\cite{hu2024routerbench}. While these methods are practical for certain tasks, they have limited applicability in scenarios where APIs often interact with user data and remote execution is unsuitable due to privacy concerns~\cite{koh2024visualwebarena}. Last, tool- or query-caching~\cite{singh2024llmcache,hu2024memserve} has been explored for efficient function calling, based on the intuition of storing and reusing recently or frequently used LLM outputs. While these methods offer latency improvements, their effectiveness applies mainly to cloud-based environments where storage is not constrained. 

Moreover, ToolLLM~\cite{qin2024toolllm} employs a tree-based scheme to minimize the number of tools required for task execution, yet it requires LLM calls across the entire tool set, making it impractical for edge devices -- where delay and power consumption are critical -- while still suffering from the same limitations as the ``off-the shelf'' LLM baselines, such as limited window sizes. In~\cite{patil2023gorilla}, the authors employ a RAG-based tool search scheme to identify relevant tools, while TinyAgent~\cite{erdogan2024tinyagent} uses a transformer-based classifier for tool selection. Octopus~\cite{chen2024octopus} improves function calling by fine-tuning smaller LLMs with token-masking to address response misalignment. However, while these approaches enhance performance for the specific datasets they are trained on, they are not scalable across diverse function spaces, as they require extensive fine-tuning. In contrast, our method simplifies the task without the need for fine-tuning, allowing for easy adaptation to new tools.

Following OpenAI's parallel function-calling release~\cite{openai2024fun}, tool ``compilation''~\cite{kim2023llmcompiler,singh2024llmcompiler} has been explored to optimize agent performance by optimizing the number of tools executed per each LLM call. However, these methods require standalone LLM calls using full-precision GPT models to perform the compiler logic, making them impractical for edge deployments, where computational resources and power are limited~\cite{karatzas2024balancing,fore2024geckopt,karatzas2024mapformer}. In contrast, our approach uses a much simpler and lightweight similarity-based mechanism, which can be handled efficiently by the deployed LLM itself. We enable this with an inexpensive, pretrained embedding tokenizer, eliminating the need for full-precision models and reducing both execution time and power consumption significantly.

%% file: methodology.tex
\begin{figure*}[!t]
    \centering
    \vspace*{-2em}
    \resizebox{0.85\textwidth}{!}{
    \includegraphics[page=4, width=\linewidth, clip, trim={2em 4 3 3}]{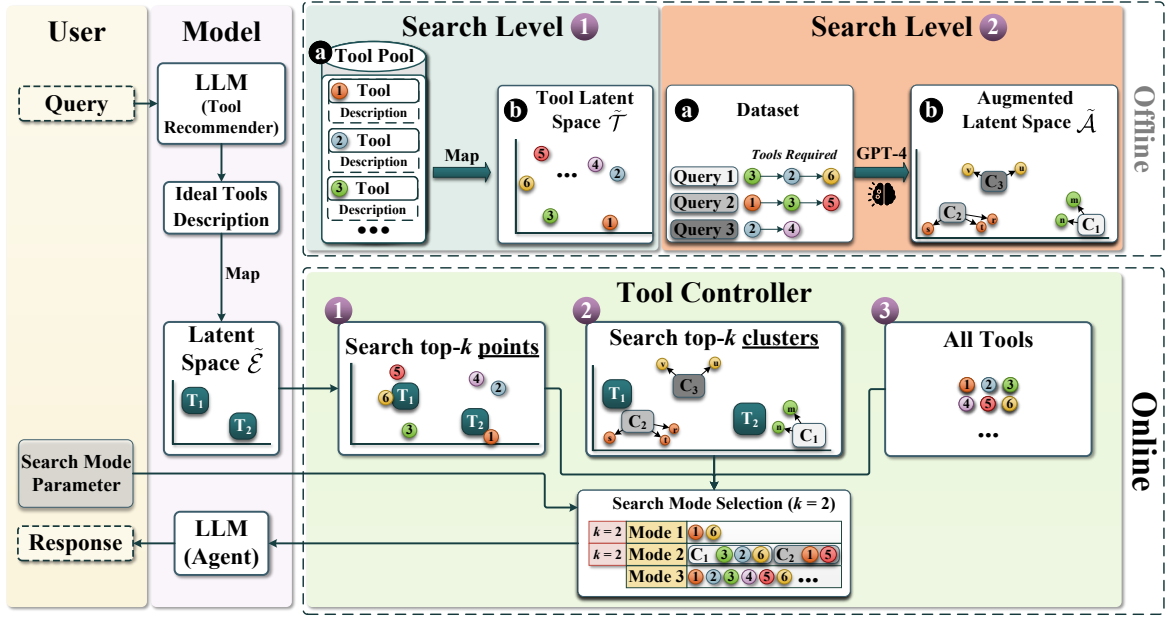}
    }
    \caption{\texttt{Less-is-More} considers three distinct tool representations (\textbf{Search Levels}): \textit{individual} tools (\textbf{Level 1}), tool \textit{clusters} (\textbf{Level 2}), or the \textit{entire} tool set (\textbf{Level 3}), whose latent spaces are constructed \textit{offline}. At runtime, given a query and \ul{\textbf{without}} providing any tools, the LLM (\textbf{Recommender}) generates tool descriptions ``ideal'' for the task. Given the LLM-recommended tool embeddings, the \textbf{Controller} identifies the most relevant \textit{real} tools based on their proximity in the latent space representations.
    }
    \label{fig:method:high-level-overview}
    \vspace*{-1.5em}
\end{figure*}

\section{Methodology}\label{sec:methodology}

\texttt{Less-is-More} simplifies LLM function calling on edge devices by dynamically reducing the number of available tools, enhancing task-completion performance, execution speed, and power efficiency, without requiring fine-tuning or complex model adjustments. Figure~\ref{fig:method:high-level-overview} provides an overview of our approach, which consists of three main components:
\begin{inparaenum}[(\bgroup\bfseries i\egroup)]
    \item a set of \textbf{Search Levels} constructed \textit{offline} with varying tool-description granularity, 
    \item a \textbf{Tool Recommender} which \textit{at runtime} and for each user prompt generates tool descriptions needed to complete the task, and
    \item a \textbf{Tool Controller} which finalizes tool selection with respect to the selected \textbf{Level}.
\end{inparaenum}

The core idea behind \texttt{Less-is-More} is to \textit{avoid presenting the LLM with all tools upfront}. \ul{\textbf{Without}} providing any APIs at first, the LLM is instead instructed to reason about the number and type of tools it ``believes'' it requires to answer the user's query, with the \textbf{Recommender} generating ``ideal'' tool definitions based on its reasoning. Based on these LLM-generated descriptions, we employ ``similarity search'' to identify the actual tools that are ``closest'' to the LLM's suggestions, with the \textbf{Controller} determining the appropriate \textbf{Search Level}.

\subsection{Constructing the \textbf{Search Levels} offline}

To address varying levels of query complexity and performance demands, we consider three distinct \textbf{Search Levels}: either selecting from \textit{individual} tools directly (\textbf{Search Level 1}), from \textit{clusters} of tools (\textbf{Search Level 2}), or the \textit{entire} tool set (\textbf{Search Level 3}). In constructing these search spaces, we aim to minimize the overhead induced at runtime for tool selection. Drawing inspiration from RAG-based similarity search~\cite{anthropic-rag}, we map all tool descriptions into embeddings, building a 768-dimensional latent space using a pre-trained embedding model based on the MPNet architecture~\cite{song2020mpnet}. 

Akin to RAG vectorstores~\cite{anthropic-rag}, this embedding encodes textual API descriptions into numerical vectors that capture their semantic meaning with two key advantages:
\begin{inparaenum}[(\bgroup\bfseries i\egroup)]
    \item for a given set of API tools, it requires only an \textbf{one-time} offline step to populate the search levels, placing tools with similar functions or characteristics close to each other in the latent space. Moreover
    \item at runtime, this spatial arrangement allows for efficient similarity measurements of tools to user queries, since the same tokenization can be applied to the LLM-generated ``ideal'' tool descriptions and match them against the stored ``real'' APIs.
\end{inparaenum}
With this representation, \texttt{Less-is-More} can quickly identify which tools are most relevant to the prompted query based on their proximity in the latent space. 

\vspace{+1pt}
\noindent
\textbf{Search Level 1 - Individual tools}: The latent space, denoted as $\tilde{\mathcal{T}}$, consists of embeddings derived from the descriptions of all available API tools using the MPNet model. This mapping of tool descriptions into the latent space $\tilde{\mathcal{T}}$ is performed as a preprocessing step, offline and prior to any user interaction, ensuring efficient real-time query processing. At this finest tool granularity, \textbf{Level 1} prioritizes speed as it enables matching LLM-generated tool descriptions directly to individual tools for straightforward queries (e.g., single-function tasks). However, since this level relies only on the latent space closeness of tools, it might compromise accuracy for more complex queries that require interactions between multiple functions. To this end, we introduce a more complex search level, %to address the limitations of \textit{standalone} tool-matching
as discussed next.

\vspace{+1pt}
\noindent
\textbf{Search Level 2 - Tool clusters}: Our goal is to construct a coarser latent space with clusters of tools. However, a clustering algorithm based on tool (text) descriptions would produce groups that poorly capture tool-usage patterns. For example, a task to translate a document and then open it in a browser would require document-related and UI-related tools, which rudimentary clustering will place in separate groups. 

To this end, we draw inspiration from benchmark \textit{augmentation} methods~\cite{zhuang2023toolqa,koh2024visualwebarena}. In these approaches, GPT is first used to augment an existing dataset by generating queries contextually similar to those in the original pool. Subsequently, GPT generates additional queries containing tasks that are contextually proximate to the tasks present in the original queries. As both GeoEngine~\cite{singh2024geollm} and BFCL~\cite{berkeley-function-calling-leaderboard} provide a categorization of their benchmark question types (e.g., wiki questions, document tasks, math tools, etc.), we randomly sample 10 queries per category from their \ul{\textit{training}} sets. We then follow ToolQA~\cite{zhuang2023toolqa} and we prompt GPT-4 \texttt{(Turbo 0125)} to generate queries with contextually proximate tasks and their respective solutions. For instance, in the previous example, where the original query involved opening a document, a task permutation might involve printing it instead. Note here that factual ``correctness'' for these generated queries is less critical, as they are not used for training or runtime decisions~\cite{zhuang2023toolqa}. Instead, they only serve as ``noisy'' queries for clustering, and we measure their quality based on a similarity score (i.e., ROUGE score following ~\cite{zhuang2023toolqa,singh2024geollm}), ensuring diverse tool combinations without redundancy. 

We map these newly generated queries into a latent space using the MPNet model, thereby augmenting our initial representation. We denote this \enquote{augmented latent space} as $\tilde{\mathcal{A}}$. We then apply Agglomerative Clustering~\cite{scikit}, i.e., a recursively clustering algorithm which starts by treating each query as its own cluster and then progressively merges the most similar clusters based on their similarity-distance in $\tilde{\mathcal{A}}$. With this ``augmented'' clustering performed offline, the coarser tool groups in this \textbf{Search Level 2} capture synergistic relationships between tools prevalent in the underlying benchmarks.

\vspace{+1pt}
\noindent
\textbf{Search Level 3 - Entire Tool Set}: This corresponds to the default function-calling logic where the agent receives \textit{all} APIs. We represent this directly with the complete tool set (text JSON format in LLM calls) without the need for a latent space, since no tool search is performed.

\subsection{Tool \textbf{Recommender} and query latent space embedding}

At runtime, given the user query and \ul{\textbf{without}} providing any tools, the LLM (tool recommender) is prompted to generate descriptions of the ``ideal'' tools it believes would be necessary to complete the task. The LLM returns a structured response in a JSON format containing the functionality of each recommended tool in detail. For example, if a user prompts, ``What's the weather like in New York and can you translate that information into French?'' the LLM would generate definitions of a ``weather\_information()'' and a ``text\_translation()'' tool, explaining that one fetches weather data and the other translates text. Since no tools are executed yet (i.e., no tool APIs are appended to the input prompt), this \textbf{Recommender} step introduces negligible overhead compared to the subsequent function calling, as shown in our Results (Section~\ref{sec:evaluation}). Using the same pretrained MPNet tokenizer, these LLM-recommended tool descriptions, alongside the corresponding user task, are mapped into 728-dimensional embeddings, denoted as $\tilde{\mathcal{E}}$, enabling efficient similarity tool-matching as described in the next step.

\subsection{Tool \textbf{Controller}}
 
Given the LLM-recommended query-tools embedding $\tilde{\mathcal{E}}$ and the tool space representations ($\tilde{\mathcal{T}}$ and $\tilde{\mathcal{A}}$), we efficiently identify the most relevant tools based on their proximity in the latent space. Following well-established RAG-based principles, our \textbf{Controller} runs a $k$-Nearest Neighbors (k-NN) search using FAISS similarity~\cite{douze2024faiss} against both \textbf{Search Level 1} and \textbf{Level 2}, retrieving the top-$k$ tools -- both individual tools and clusters -- that closely match the LLM's tool-selection reasoning. We then compare the top-$k$ similarity values and select the \textbf{Search Level} with the highest average score. 

Intuitively, for simpler queries that require only a single tool (where the LLM would likely recommend just one ``ideal'' tool description), similarity scores would be higher against standalone tool descriptions. In contrast, LLM recommendations involving multiple tools are more likely to match a tool cluster. Finally, we proceed with function calling, invoking the LLM with only the subset of tools identified by the \textbf{Controller}.

We note that prior work has leveraged similarity-based tool selection~\cite{patil2023gorilla, qin2024toolllm}, but their search is conducted against the entire tool ontology -- which closely resembles running only \textbf{Level 1} in our method. In contrast, to our knowledge, we are the first to consider varying levels of tool embedding granularity without resorting to complex tree-based search schemes, enabling function calling that balances task complexity, tool diversity, and edge hardware overhead trade-offs.

To handle LLM errors at runtime, we employ a ``fallback'' mechanism, as is common in LLM ``compilers''~\cite{kim2023llmcompiler}. In our prompt, we also instruct the LLM to signal a failure by returning an error message if the function-calling step fails after retrying. If an error occurs, the next tool-calling attempt defaults to ``vanilla'' function-calling with \textit{all} APIs provided (\textbf{Level 3}). Similarly, to account for any \textbf{Recommender} mistakes, if both average top-$k$ scores are below $0.5$, indicating low confidence in selecting either Level 1 or Level 2, we default to presenting all tools (Level 3) to the LLM. 

%% file: evaluation.tex
\begin{figure*}[!htpb]
    \centering
    \footnotesize 
    \resizebox{0.98\textwidth}{!}{    
    \includegraphics[page=4, width=\linewidth, clip, trim={0em 0 0 0}]{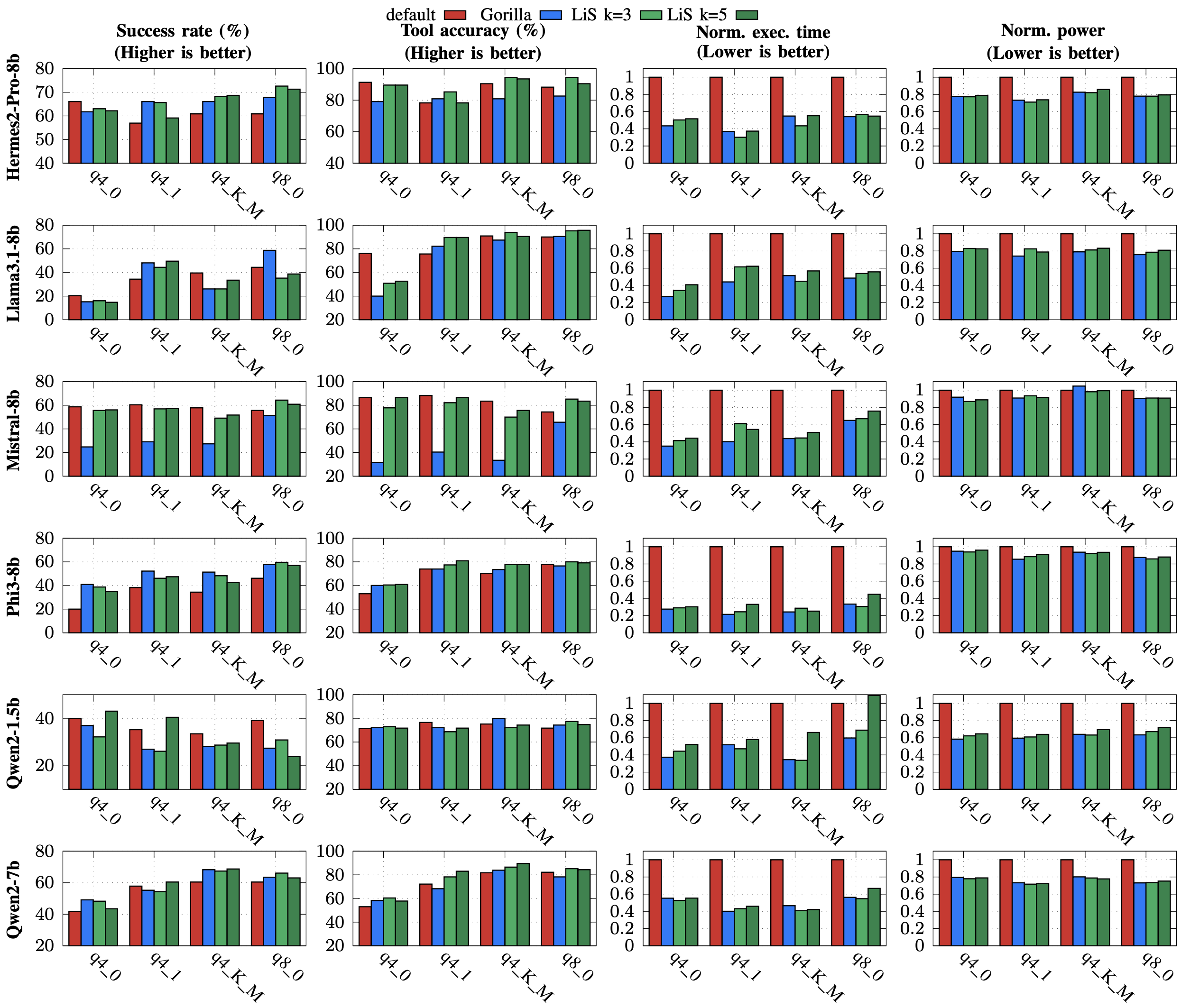}
    }
    \vspace{-2pt}
    \caption{Performance comparison of our method (LiS) at varying $k$ values against the default approach, measured by Success Rate, Tool Accuracy, Normalized Execution Time, and Normalized Power for the BFCL~\cite{berkeley-function-calling-leaderboard} benchmark.}
    \label{fig:gorilla_resutls}
    \vspace{-12pt}
\end{figure*}

\section{Experimental Results}\label{sec:evaluation}

We evaluate our method using two benchmarks: the BFCL benchmark~\cite{berkeley-function-calling-leaderboard} for general function calling and GeoEngine~\cite{singh2024geollm} for application-specific function calling. BFCL~\cite{berkeley-function-calling-leaderboard} mainly involves single function calls for each query, even when the query contains unrelated sub-questions, while GeoEngine focuses on geographic applications requiring sequential function calls, where each call depends on the previous result~\cite{singh2024evaluating}. In contrast to GeoEngine, BFCL is simpler because it handles each sub-question independently, without needing to process information sequentially across multiple function calls. Interestingly, we found that in BFCL \textbf{Search Level 1} yields higher tool-matching scores, whereas  
for GeoEngine it is \textbf{Search Level 2} with better tool selection. We conducted our experiments using mini-batches of 230 queries from each benchmark, along with 51 functions from BFCL and 46 functions from GeoEngine. 

\begin{figure*}
    \centering
    \footnotesize 
    \resizebox{0.98\textwidth}{!}{    
    \includegraphics[page=4, width=\linewidth, clip, trim={0em 0 0 0}]{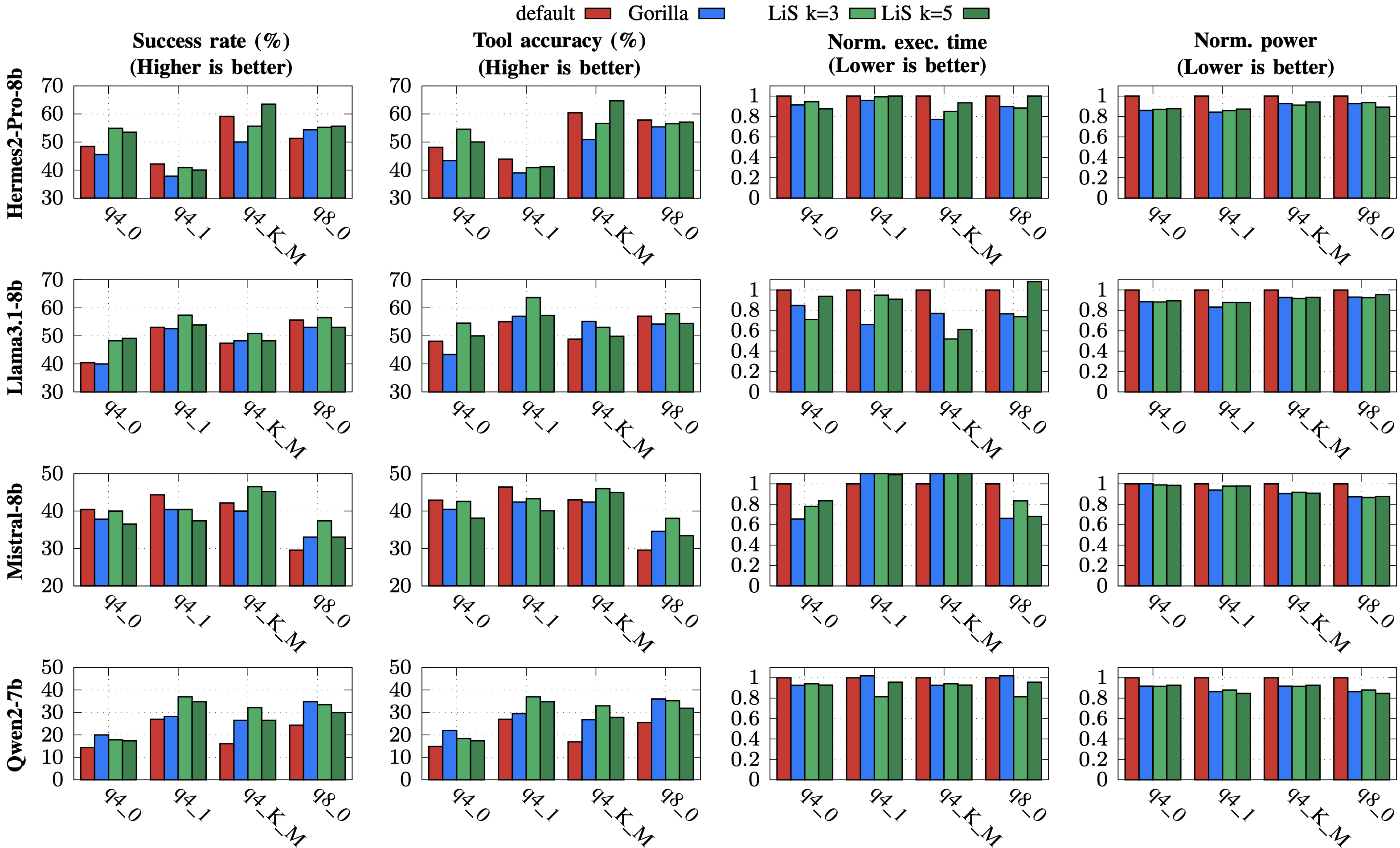}
    }
    % \vspace{2pt}
    \caption{Performance comparison of our method (LiS) at varying $k$ values against the default approach, measured by Success Rate, Tool Accuracy, Normalized Execution Time, and Normalized Power for the GeoEngine~\cite{singh2024geollm} benchmark.}
    \vspace{-10pt}
    \label{fig:geoengine_resutls}
\end{figure*}

We used the NVIDIA AGX Orin~\cite{karumbunathan2022nvidia} board as our edge device.
% This board features a 2048-core GPU based on the NVIDIA Ampere architecture with 64 Tensor Cores, a 12-core Arm Cortex-A78 CPU, and 32GB of memory.
We focused on four key metrics: \begin{inparaenum}[(i)]
\item \ul{Tool Accuracy}~\cite{patil2023gorilla}: the frequency with which the LLM selected the correct tool from the available options,
\item \ul{Success Rate}~\cite{singh2024geollm}: it shows whether the LLM not only chose the correct tool but also used it properly, such as providing the correct input types according to the function's requirements;
\item \ul{Normalized Execution Time:} the average time taken to complete a query, adjusted relative to the baseline, and
\item \ul{Normalized power consumption:} the average power used by the device for processing each query, adjusted relative to the baseline.
\end{inparaenum}

The LLMs we evaluated are: \begin{inparaenum}[(i)]
    \item \ul{Hermes2-Pro-8b}~\cite{hermes}, an advanced LLaMA variant optimized for natural language understanding and function calling;
    \item \ul{Llama3.1-8b}~\cite{dubey2024llama}, a state-of-the-art model fine-tuned for language tasks;
    \item \ul{Mistral-8b}~\cite{jiang2023mistral}, known for speed and efficiency;
    \item \ul{Phi3-8b}~\cite{abdin2024phi}, specialized for specific tasks;
    \item \ul{Qwen2-1.5b}~\cite{yang2024qwen2}, a smaller model optimized for performance on limited resources; and
    \item \ul{Qwen2-7b}, a larger version for handling more complex language tasks.
\end{inparaenum}
For each model, we evaluated four quantization variants: \texttt{q4\_0} (4-bit quantization for memory efficiency), \texttt{q4\_1} (an optimized version with improved accuracy), \texttt{q4\_K\_M} (introducing mixed-precision for balanced performance), and \texttt{q8\_0} (8-bit quantization for higher precision but increased memory use). For \texttt{Less-is-More} (LiS), we tested with $k=3$, and $k=5$. We compared our method to the default execution, where LLMs access all tools, and to Gorilla~\cite{patil2023gorilla}, which uses similarity-based methods to identify the most likely tool. We also attempted to compare against ToolLLM~\cite{qin2024toolllm}, but its tree-based exploration could not fit on the board.

We determined the minimum context window required for each model to fit all tools and handle communication with the LLM, both with and without \texttt{Less-is-More}. For default models used via Ollama~\cite{ollama-tool}, the context window was set to $16k$, while Gorilla and \texttt{Less-is-More} and reduced this to $8k$ across all $k$ values. For the default models, we also tested context windows larger than $16k$. While there was no significant improvement in success rate, execution time increased noticeably, which is why we chose the $16k$ value.

Figure~\ref{fig:gorilla_resutls} shows the results for the BFCL benchmark. \textbf{Hermes2-Pro-8b}: Compared to the default model, the optimized versions achieved a significant increase to approximately 71\% in success rate. Gorilla was also better than the default, but lower than LiS. LiS also improved tool accuracy to 89\%, reflecting a better ability to select and use tools correctly. Notably, execution time was reduced by up to 80\% on the device, and power consumption decreased by 45\% at best, making this model much more efficient for edge deployment. \textbf{Llama3.1-8b:} The success rate increased from the baseline to 44.2\% and tool accuracy reached 93.8\%. These optimizations led to a 72\% reduction in execution time and a 30\% decrease in power consumption. \textbf{Mistral-8b}: For Mistral-8b, even though the optimizations did not result in any gain in success rate and tool accuracy, our method resulted in a 77\% reduction in execution time and a 18\% decrease in power consumption. Gorilla was the worst in success rate and tool accuracy mainly due to the limited capabilities of compressed Mistral. \textbf{Phi3-8b}: Our method improved the success rate to 55\% and tool accuracy to 78\%. Execution time was reduced by 55\%, and power consumption decreased by 20\%, enhancing the model's overall efficiency. \textbf{Qwen2-1.5b}: Although Qwen2-1.5b is a smaller model, the optimizations led to a noticeable increase in success rate to approximately 40\% and tool accuracy to 76\%. Execution time was reduced by 48\%, and power usage dropped by 20\%, showing that even lightweight models can benefit greatly from our method. \textbf{Qwen2-7b}: The larger Qwen2-7b model achieved an increase in success rate to 68\% and tool accuracy to 87\%, remarkably better than its default performance. Execution time was reduced by up to 70\%, and power consumption improved by 27\%, enabling this model to handle more complex tasks efficiently on edge devices. Overall, our method improved all four metrics—success rate, tool accuracy, execution time, and power consumption—across all the LLMs. The improvements in success rate and tool accuracy were achieved by reducing the number of tools available to the LLM, which minimized confusion and allowed it to make better decisions. Additionally, by speeding up the LLM's decision-making process and using a smaller context window, we further reduced execution time and achieved significant gains in power efficiency.

Figure~\ref{fig:geoengine_resutls} depicts the results on the GeoEngine benchmark which is more complex than BFCL.
Given this complexity, the Phi3 and Qwen2-1.5b models exhibited a low default success rate of approximately 10\%, making their power consumption and execution time measurements unreliable. To maintain fairness, we have excluded these two LLMs from our analysis.
\textbf{Hermes2-Pro-8b}:
Our optimizations improved the success rate to 63\% and enhanced tool accuracy to 64\%. Additionally, we achieved a 15\% reduction in execution time and a 6\% decrease in power consumption. \textbf{Llama3.1-8b}: The success rate improved to 56\% similarly to tool accuracy. Execution time was reduced by up to 40\%, and power consumption decreased by approximately 12\%, significantly enhancing the model's efficiency for edge devices. \textbf{Mistral-8b}: The optimizations resulted in a success rate increase to 46\%, with tool accuracy improving to 47\%. Execution time was 10\% higher for some variations, compensated by the big increase in success rate. Power consumption decreased by 9\%, making Mistral-8b more power-efficient. \textbf{Qwen2-7b}:  The success rate improved to 35\%, with tool accuracy reflecting a similar trend. The optimizations resulted in a 21\% reduction in execution time and a notable decrease in power consumption of about 13\%. Overall, Gorilla struggled to improve the success rate in most cases as it only checks tool similarity, while GeoEngine requires sequential function calls where each depends on the previous result.

%% file: conclusion.tex
% \vspace{-5pt}
\section{Conclusion}\label{sec:conclusion}
% \vspace{-5pt}
In this paper, we introduced \texttt{Less-is-More}, a novel method for optimizing function calling in Large Language Models (LLMs) deployed on edge devices. By reducing the number of available tools and using hierarchical search levels, our method improves success rate, execution speed, and power efficiency without the need for extensive fine-tuning. Experimental results demonstrate significant improvements in success rate, tool accuracy, execution time, and power consumption compared to default approaches, particularly for complex queries. Overall, our approach provides a practical solution for deploying LLMs on resource-constrained devices, opening new possibilities for efficient and scalable edge AI applications.

%% file: acknowledgment.tex
\section*{Acknowledgments}

This work is supported by grant NSF CCF 2324854.